\documentstyle[12pt,fleqn,rotate,epsfig]{article}
\newcommand{\be}{\begin{equation}}
\newcommand{\ee}{\end{equation}}
\newcommand{\bea}{\begin{eqnarray}}
\newcommand{\eea}{\end{eqnarray}}

\newcommand{\php}{\phantom{'}\!\!}

\renewcommand{\appendix}{
  \setcounter{section}{0}
  \renewcommand{\thesection}{\Alph{section}}
}

\begin{document}
\title{Equilibration within a semiclassical off-shell transport
approach\footnote{supported by GSI Darmstadt}}
\author{W. Cassing and S. Juchem\\
Institut f\"ur Theoretische Physik, Universit\"at Giessen\\
35392 Giessen, Germany}
\maketitle
\begin{abstract}
Equilibration times for nuclear matter configurations -- modelling
intermediate and high energy nucleus-nucleus collisions -- are
evaluated within the semiclassical off-shell transport approach
developed recently. The transport equations are solved for a
finite box in coordinate space employing periodic boundary
conditions. The off-shell transport model is shown to give proper
off-shell equilibrium distributions in the limit $t \rightarrow
\infty$ for the nucleon and $\Delta$-resonance spectral functions.
We find that equilibration times within the off-shell approach are
only slightly enhanced as compared to the on-shell limit for the
momentum configurations considered.
\end{abstract}

\vspace{2cm}

\newpage
\section{Introduction}
The dynamical description of strongly interacting systems out of
equilibrium nowadays is dominantly based on transport theories and
efficient numerical recipies have been set up for the solution of
the coupled channel transport equations \cite{URQMD,CB99} (and
Refs. therein). These transport approaches have been derived
either from the Kadanoff-Baym equations \cite{kb62} in Refs.
\cite{pd841,Bot,Mal,ph95,gl98} or from the hierarchy of connected
equal-time Green functions \cite{Zuo,Wang1} in Refs.
\cite{CaWa,CNW} by applying a Wigner transformation and
restricting to first order in the derivatives of the phase-space
variables ($X, P$). Whereas theoretical formulations of off-shell
quantum transport have been limited to the formal level for a
couple of years \cite{Bot,ph95,Rudy1,Rudy2} only recently a
tractable semiclassic form has been derived for testparticles in
the eight dimensional phase space of a particle
\cite{ca1,ca2,Leupold}.

 Whereas in Refs. \cite{ca1,ca2} we have
investigated the off-shell transport approach with respect to
nucleus-nucleus collisions at GANIL, SIS and AGS energies, we here
concentrate on equilibration phenomena relative to the on-shell
dynamics by imposing periodic boundary conditions for the system
confined to a box of size $V= L^3$, where $L$ denotes the length
of the cubic box. We, furthermore, compare the equilibrium nucleon
and $\Delta$ distribution functions ($t \rightarrow \infty$)
to the statistical model (SM)
employing the same spectral functions as in the transport
approach. For related studies at higher bombarding energies or
energy densities within on-shell transport approaches we refer the
reader to Refs.  \cite{Brat2000,Brav1,Brav2,Brav3,Solfr99}.

\section{Extended semiclassical transport equations}
We briefly recall the basic equations for Green functions and
particle self energies  that are exploited in the derivation of
off-shell transport equations in the semiclassical limit.

The general starting point for the derivation of a transport
equation for particles with a finite and dynamical width are the
Dyson-Schwinger equations for the retarded and advanced Green
functions $S^{ret}$, $S^{adv}$ and for the non-ordered Green
functions $S^{<}$ and $S^{>}$ \cite{kb62}. In the case of scalar
bosons -- which is considered in the following for simplicity --
these Green functions are defined by
\bea \hspace{2.0cm} i \, S^{<}_{xy} & := & \; < \,
\Phi^{\dagger}(y) \; \Phi(x) \, > \, ,  \; \; i \, S^{>}_{xy}  :=  \;
<\,  \Phi(x) \; \Phi^{\dagger}(y) \, > \, , \nonumber\\ i \,
S^{ret}_{xy} & := & \phantom{-} \: \Theta (x_0 - y_0) \, < \, [ \,
\Phi(x) \, , \, \Phi^{\dagger}(y) \, ] \, > \, , \nonumber\\ i \,
S^{adv}_{xy} & := & - \: \Theta (y_0 - x_0) \, < \, [ \, \Phi(x)
\, , \, \Phi^{\dagger}(y) \, ] \, > \, . \eea
They depend on the space-time coordinates $x,y$ as indicated by
the indices $\cdot_{xy}$. The Green functions are determined via
Dyson-Schwinger equations by the retarded and advanced self
energies $\Sigma^{ret},\Sigma^{adv}$ and the collisional self
energy $\Sigma^{<}$:
\bea \hat{S}_{0x}^{-1} \; S_{xy}^{ret} \; \: = \; \: \delta_{xy}
\; \: + \; \: \Sigma_{xz}^{ret} \: \odot \: S_{zy}^{ret} \; ,
\label{dsret_spatial} \eea
\bea \hat{S}_{0x}^{-1} \; S_{xy}^{adv} \; \: = \; \: \delta_{xy}
\; \: + \; \: \Sigma_{xz}^{adv} \: \odot \: S_{zy}^{adv} \; ,
\label{dsadv_spatial} \eea
\bea \hat{S}_{0x}^{-1} \; S_{xy}^{<} \; \: = \; \:
\Sigma_{xz}^{ret} \: \odot \: S_{zy}^{<} \; \: + \; \:
\Sigma_{xz}^{<} \: \odot \: S_{zy}^{adv} \: , \label{kb_spatial}
\eea \\
where Eq. (\ref{kb_spatial}) is the well-known Kadanoff-Baym
equation. Here $\hat{S}^{-1}_{0x}$ denotes the (negative)
Klein-Gordon differential operator which is given for bosonic
field quanta of (bare) mass $M_0$ by $\hat{S}^{-1}_{0x} = -
(\partial^{\mu}_{x} \partial^{x}_{\mu} + M^2_0 )$; $\delta_{xy}$
represents the four-dimensional $\delta$-distribution $\delta_{xy}
\equiv \delta^{(4)}(x-y)$ and the symbol $\odot$ indicates an
integration (from $-\infty$ to $\infty$) as well as a summation
over all discrete intermediate variables (cf. \cite{ph95,ca1}).

\subsection{The semiclassical limit}
For the derivation of a semiclassical transport equation one now
changes from a pure space-time formulation into the
Wigner-representation. The theory is then formulated in terms of
the center-of-mass variable $X = (x+y)/2$ and the momentum $P$,
which is introduced by Fourier-transformation with respect to the
relative space-time coordinate $(x-y)$. In any semiclassical
transport theory one, furthermore, keeps only contributions up to
the first order in the space-time gradients. After carrying-out
these two steps the Dyson-Schwinger equations
(\ref{dsret_spatial})-(\ref{kb_spatial}) become
\bea \left[ \, P^2 \, - \, M^2_0 \, + \, i P^{\mu}
\partial^{X}_{\mu} \, \right] \: S^{ret}_{XP} \; = \; 1 \: + \: (
\, 1 \, - \, i \, \Diamond \, ) \, \{ \, \Sigma^{ret}_{XP} \, \}
\: \{ \, S^{ret}_{XP} \, \} \: , \label{dsret_wignerfo} \eea
\bea \left[ \, P^2 \, - \, M^2_0 \, + \, i P^{\mu}
\partial^{X}_{\mu} \, \right] \: S^{adv}_{XP} \; = \; 1 \: + \: (
\, 1 \, - \, i \, \Diamond \, ) \, \{ \, \Sigma^{adv}_{XP} \, \}
\: \{ \, S^{adv}_{XP} \, \} \: , \label{dsadv_wignerfo} \eea
\bea \left[ \, P^2 \, - \, M^2_0 \, + \, i P^{\mu}
\partial^{X}_{\mu} \, \right] \: S^{<}_{XP} \; = \; ( \, 1 \, - \,
i \, \Diamond \, ) \, \left[ \: \{ \, \Sigma^{ret}_{XP} \, \} \:
\{ \, S^{<}_{XP} \, \} \; + \;
% ( \, 1 \, - \, i \, \Diamond \, ) \,
\{ \, \Sigma^{<}_{XP} \, \} \: \{ \, S^{adv}_{XP} \, \} \: \right]
, \label{kb_wignerfo} \eea\\
where the operator $\Diamond$ is defined as \cite{ph95,ca1}
\bea
 \Diamond \, \{ \, F_{1} \, \} \, \{ \, F_{2} \, \}
\; := \; \frac{1}{2} \left( \frac{\partial F_{1}}{\partial
X^{\mu}} \: \frac{\partial F_{2}}{\partial P_{\mu}} \; - \;
\frac{\partial F_{1}}{\partial P_{\mu}} \: \frac{\partial
F_{2}}{\partial X^{\mu}} \right) .\label{poissonoperator} \eea\\
It is a four-dimensional generalization of the well-known
Poisson-bracket. Starting from (\ref{dsret_wignerfo}) and
(\ref{dsadv_wignerfo}) one obtains algebraic relations between the
real and the imaginary part of the retarded Green functions. On
the other hand eq. (\ref{kb_wignerfo}) leads to a 'transport
equation' for the Green function $S^{<}$ \cite{ca1}.

To this aim one separates all retarded and advanced quantities --
Green functions and self energies  -- into real and imaginary
parts,
\bea S_{XP}^{ret,adv} \; = \;
%G_{XP}
Re S^{ret}_{XP} \; \mp \; \frac{i}{2} \, A_{XP} \; , \qquad
\Sigma_{XP}^{ret,adv} \; = \; Re \Sigma^{ret}_{XP} \; \mp \;
\frac{i}{2} \, \Gamma_{XP} \; . \label{ret_sep} \eea\\
The imaginary part of the retarded propagator is given (up to a
factor 2) by the normalized spectral function
\bea A_{XP} \: = \: i \left[ \, S_{XP}^{ret} \: - \: S_{XP}^{adv}
\, \right] \: = \: - 2 \, Im \, S^{ret}_{XP} \; , \qquad \qquad
\qquad \int \frac{d P_0^2}{4 \pi} \,  A_{XP} \; = \; 1 \; ,
\label{spectralfunction} \eea\\
while the imaginary part of the self energy corresponds to half
the particle width $\Gamma_{XP}$. By separating the complex equations
(\ref{dsret_wignerfo}) and (\ref{dsadv_wignerfo}) into their real
and imaginary contributions we obtain an algebraic equation
between the real and the imaginary part of $S^{ret}$,
\bea Re S^{ret}_{XP} \; = \; \frac{P^{2} \: - \: M_{0}^{2} \: - \:
Re \Sigma^{ret}_{XP}}{\Gamma_{XP}} \; A_{XP} \, .
\label{dispersion} \eea\\
In addition we gain an algebraic solution for the spectral
function as
\bea A_{XP} \; = \; \frac{ \Gamma_{XP} } {( \, P^2 \, - \,
M_{0}^{2} \, - \, Re \Sigma^{ret}_{XP} )^{2} \: + \:
\Gamma_{XP}^{2}/4} \; , \label{alg_spectral} \eea\\
while the real part of the retarded propagator is given by
\bea Re S^{ret}_{XP} \; = \; \frac{P^{2} \: - \: M_{0}^{2} \: - \:
Re \Sigma^{ret}_{XP}} {( \, P^2 \, - \, M_{0}^{2} \, - \, Re
\Sigma^{ret}_{XP} )^{2} \: + \: \Gamma_{XP}^{2}/4} \, .
\label{alg_realpart} \eea \\
Furthermore, the (Wigner-transformed) Kadanoff-Baym equation
(\ref{kb_wignerfo}) allows for the construction of a transport
equation for the Green function $S^{<}$. When separating the real
and the imaginary contribution of this equation we find i) a
generalized transport equation,
\bea \Diamond \, \{ \, P^{2} &-& M_{0}^{2} \: - \: Re
\Sigma^{ret}_{XP} \, \} \; \{ \, S^{<}_{XP} \, \} \; - \; \Diamond
\, \{ \, \Sigma^{<}_{XP} \, \} \; \{ Re S^{ret}_{XP} \, \}
\nonumber\\[0.2cm] &=& \frac{i}{2} \: \left[ \: \Sigma^{>}_{XP} \:
S^{<}_{XP} \; - \;
          \Sigma^{<}_{XP} \: S^{>}_{XP} \: \right],
\label{general_transport} \eea\\
and ii) a generalized mass-shell constraint
\bea [ \, P^{2} &-& M_{0}^{2} \: - \: Re \Sigma^{ret}_{XP} \, ] \;
S^{<}_{XP} \; - \; \Sigma^{<}_{XP} \; Re S^{ret}_{XP}
\nonumber\\[0.2cm] &=& \frac{1}{2} \, \Diamond \; \{ \,
\Sigma^{<}_{XP} \, \} \; \{ \, A_{XP} \, \} \; - \; \frac{1}{2} \,
\Diamond \; \{ \, \Gamma_{XP} \, \} \; \{ \, S^{<}_{XP} \, \} \; .
\label{general_massshell} \eea\\
Besides the drift term (i.e. $\Diamond \{P^2 - M^2_0\} \{ S^{<} \}
= - P^{\mu} \partial^{X}_{\mu} S^{<} )$ and the Vlasov term (i.e.
$-\Diamond \{ Re \Sigma^{ret} \} \{ S^{<} \} $) a third
contribution appears on the l.h.s. of (\ref{general_transport})
(i.e. $-\Diamond \{\Sigma^{<} \} \{Re S^{ret} \}$), which vanishes
in the quasiparticle limit and incorporates -- as shown in
\cite{ca1,ca2,Leupold} -- the off-shell behaviour in the particle
propagation which has been neglected so far in transport
studies\footnote{This also holds true for the recent numerical
off-shell simulations  in Refs. \cite{Effe1,Effe}}. The r.h.s. of
(\ref{general_transport}) consists of a collision term with its
characteristic gain ($\sim \Sigma^{<} S^{>}$) and loss ($\sim
\Sigma^{>} S^{<}$) structure, where scattering and decay processes of
particles  are described.

Within the specific term ($\Diamond \{\Sigma^{<} \} \{Re S^{ret}
\}$) a further modification is necessary. According to Botermans
and Malfliet \cite{Bot} the collisional self energy $\Sigma^{<}$
should be replaced by $S^{<} \cdot \Gamma / A$ to gain a
consistent first order gradient expansion scheme. The replacement
is allowed since the difference between these two expressions
$(\Sigma^< - S^< \cdot \Gamma/A)$ can be shown to be of first
order in the space-time gradients itself \cite{ca1}. Furthermore,
this substitution is required to get rid of the inequivalence
between the general transport equation (\ref{general_transport})
and the general mass shell constraint (\ref{general_massshell})
\cite{Knoll}.

Finally, the general transport equation (in first order gradient
expansion) reads \cite{ca1,ca2,Leupold}
\setlength{\mathindent}{-0.5cm} \bea A_{XP} \, \Gamma_{XP} &&
\!\!\!\!\! \!\!\!\!\! \left[ \, \Diamond \; \{ \, P^2 - M_0^2 - Re
\Sigma^{ret}_{XP} \, \} \; \{ \, S^<_{XP} \, \} \: - \:
\frac{1}{\Gamma_{XP}} \; \Diamond \; \{ \, \Gamma_{XP} \, \} \; \{
\, ( \, P^2 - M_0^2 - Re \Sigma^{ret}_{XP} \, ) \, S^<_{XP} \, \}
\, \right] \nonumber\\[0.2cm] && \; = \; i \, \left[ \,
\Sigma^>_{XP} \: S^<_{XP} \: - \: \Sigma^<_{XP} \: S^>_{XP} \,
\right]. \label{trans_approx} \eea \\
\setlength{\mathindent}{0.5cm}
Its formal structure is fixed by the approximations applied,
however, its physical contents is fully determined by the
different self energies, i.e. $Re \Sigma_{XP}^{ret}, \Gamma_{XP},
\Sigma_{XP}^<$ that have to be specified in
addition.

In order to obtain an approximate solution to the transport
equation (\ref{trans_approx})  a testparticle ansatz is used for
the Green function $S^{<}$, more specifically for the real and
positive semidefinite quantity $F_{XP} \; = A_{XP} N_{XP} = \; i
\, S^{<}_{XP}$,
\bea F_{XP} \;  \sim \; \sum_{i=1}^{N} \; \delta^{(3)} ({\vec{X}}
\, - \, {\vec{X}}_i (t)) \; \; \delta^{(3)} ({\vec{P}} \, - \,
{\vec{P}}_i (t)) \; \; \delta(P_0 - \, \epsilon_i(t)) \: .
\label{testparticle} \eea
In the most general case (where the self energies depend
on four-momentum $P$, time $t = X_0$ and the spatial coordinates
$\vec{X}$) the equations of motion for the testparticles  read \cite{ca2}
\bea \label{eomr} \frac{d {\vec X}_i}{dt} \! & = & \! \phantom{- }
\frac{1}{1 - C_{(i)}} \, \frac{1}{2 \epsilon_i} \: \left[ \, 2 \,
{\vec P}_i \, + \, {\vec \nabla}_{P_i} \, Re \Sigma^{ret}_{(i)} \,
+ \, \frac{ \epsilon_i^2 - {\vec P}_i^2 - M_0^2 - Re
\Sigma^{ret}_{(i)}}{\Gamma_{(i)}} \: {\vec \nabla}_{P_i} \,
\Gamma_{(i)} \: \right],
\\[0.3cm]
\label{eomp} \frac{d {\vec P}_i}{d t} \! & = & \! -
\frac{1}{1-C_{(i)}} \, \frac{1}{2 \epsilon_{i}} \: \left[ {\vec
\nabla}_{X_i} \, Re \Sigma^{ret}_i \: + \: \frac{\epsilon_i^2 -
{\vec P}_i^2 - M_0^{2} - Re \Sigma^{ret}_{(i)}}{\Gamma_{(i)}} \:
{\vec \nabla}_{X_i} \, \Gamma_{(i)} \: \right],
\\[0.3cm]
\label{eome} \frac{d \epsilon_i}{d t} \!
\setlength{\mathindent}{-0.5cm} & = & \! \phantom{- } \frac{1}{1 -
C_{(i)}} \, \frac{1}{2 \epsilon_i} \: \left[ \frac{\partial Re
\Sigma^{ret}_{(i)}}{\partial t} \: + \: \frac{\epsilon_i^2 - {\vec
P}_i^2 - M_0^{2} - Re \Sigma^{ret}_{(i)}}{\Gamma_{(i)}} \:
\frac{\partial \Gamma_{(i)}}{\partial t} \right], \eea\\
where the notation $F_{(i)}$ implies that the function is taken at
the coordinates of the testparticle, i.e. $F_{(i)} \equiv
F(t,\vec{X}_{i}(t),\vec{P}_{i}(t),\epsilon_{i}(t))$.

In (\ref{eomr}-\ref{eome}) a common multiplication factor
$(1-C_{(i)})^{-1}$ appears, which contains the energy derivatives
of the retarded self energy
\bea \label{correc} C_{(i)} \: = \: \frac{1}{2 \epsilon_i} \left[
\frac{\partial}{\partial \epsilon_i} \, Re \Sigma^{ret}_{(i)} \: +
\: \frac{\epsilon_i^2 - {\vec P}_i^2 - M_0^2 - Re
\Sigma^{ret}_{(i)}}{\Gamma_{(i)}} \: \frac{\partial }{\partial
\epsilon_i} \, \Gamma_{(i)} \right] \: , \eea
which yields a shift of the system time $t$ to the 'eigentime' of
particle $i$ defined by $\tilde{t}_{i} = t /(1-C_{(i)})$. The
derivatives with respect to the 'eigentime', i.e. $d \vec{X}_i / d
\tilde{t}_i$, $d \vec{P}_i / d \tilde{t}_i$ and $d \epsilon_i / d
\tilde{t}_i$ then emerge without this renormalization factor for
each testparticle $i$ when neglecting the explicit time dependence
of $C_{(i)}$ in line with the semiclassical approximation scheme.
The role and the importance of this correction factors have been
studied in Ref. \cite{ca2} for a four-momentum-dependent 'trial'
potential and we refer the reader to the latter analysis for more
details.

Following Ref. \cite{ca1} we take $M^{2} = P^2 - Re \Sigma^{ret}$
as an independent variable instead of the energy $P_0$.
Eq. (\ref{eome}) then
turns to
\bea \label{eomm} \frac{dM_i^2}{dt} \; = \; \frac{M_i^2 -
M_0^2}{\Gamma_{(i)}} \; \frac{d \Gamma_{(i)}}{dt} \eea
for the time evolution of the test-particle $i$ in the invariant
mass squared as derived in Refs. \cite{ca1,ca2}. We mention that
corresponding equations of motion have been derived by Leupold in
the nonrelativistic limit \cite{Leupold}.

\subsection{Generalized collision terms for bosons and fermions}
The collision term of the Kadanoff-Baym equation can only be
worked out in more detail by giving explicit approximations for
$\Sigma^{<}$ and $\Sigma^{>}$.   We recall the formulation and
result from Ref. \cite{ca2} that is based on Dirac-Brueckner
theory, i.e. $$ I_{coll}(X,\vec{P},M^2) = Tr_2 Tr_3 Tr_4 A(X,{\vec
P},M^2) A(X,{\vec P}_2, M_2 ^2) A(X,{\vec P}_3, M_3 ^2) A(X,{\vec
P}_4, M_4 ^2) $$ $$
%\{
|T(({\vec P},M^2) + ({\vec P}_2,M_2^2) \rightarrow ({\vec
P}_3,M_3^2) + ({\vec P}_4,M_4^2))|_{{\cal A,S}}^2 \; \;
\delta^{(4)}({P} + {P}_2 - {P}_3 - {P}_4) $$
\be
\label{Icoll} \hspace{1cm} [ \, N_{X{\vec P}_3 M_3^2} \, N_{X
{\vec P}_4 M_4^2} \, {\bar f}_{X {\vec P} M^2} \, {\bar f}_{X
{\vec P}_2 M_2^2} \: - \: N_{X{\vec P} M^2} \, N_{X {\vec P}_2
M_2^2} \, {\bar f}_{X {\vec P}_3 M_3^2} \, {\bar f}_{X {\vec P}_4
M_4^2} \, ]
%\}
\ee with
\be
\label{pauli} {\bar f}_{X {\vec P} M^2} = 1 + \eta \, N_{X {\vec
P} M^2} \ee and $\eta = \pm 1$ for bosons/fermions, respectively.
The indices ${\cal A,S}$ stand for the antisymmetric/symmetric
matrix element of the in-medium off-shell scattering amplitude $T$
in case of fermions/bosons. In Eq. (\ref{Icoll}) the trace over
particles 2,3,4 reads explicitly for fermions
\be
\label{trace} Tr_2 = \sum_{\sigma_2, \tau_2} \frac{1}{(2 \pi)^4}
\int d^3 P_2 \frac{d M^2_2}{2 \sqrt{\vec{P}^2_2+M^2_2}}, \ee where
$\sigma_2, \tau_2$ denote the spin and isospin of particle 2. In
case of bosons we have instead
\be
\label{trace2} Tr_2 = \sum_{\sigma_2, \tau_2} \frac{1}{(2 \pi)^4}
\int d^3 P_2 \frac{d P_{0,2}^2}{2}, \ee since here the spectral
function $A_B$ is normalized as
\be
\label{sb} \int \frac{d P_0^2}{4 \pi} A_B(X,P) = 1 \ee whereas for
fermions we have
\be
\label{sb1} \int \frac{d P_0}{2 \pi} A_F(X,P) = 1. \ee We mention
that the spectral function $A_F$ in case of fermions in
(\ref{Icoll}) is obtained by considering only particles of
positive energy and assuming the spectral function to be identical
for spin 'up' and 'down' states (cf. Ref. \cite{ca2}).

Neglecting the 'gain-term' in Eq. (\ref{Icoll}) one recognizes
that the collisional width $\Gamma_{coll}$ of the particle in the
rest frame is given by
\be
\label{gcoll} \Gamma_{coll}(X,\vec{P},M^2) = Tr_2 Tr_3 Tr_4 \;
%\{
|T(({\vec P},M^2) + ({\vec P}_2,M_2^2) \rightarrow ({\vec
P}_3,M_3^2) + ({\vec P}_4,M_4^2))|_{{\cal A,S}}^2 \ee $$ A(X,{\vec
P}_2,M_2^2) A(X,{\vec P}_3,M_3^2) A(X,{\vec P}_4, M_4^2) \; \;
\delta^4(P + P_2 - P_3-P_4) \ N_{X {\vec P}_2 M_2^2} \, {\bar
f}_{X {\vec P}_3 M^2_3} \, {\bar f}_{X {\vec P}_4 M^2_4} \,
%\}
, $$ where as in Eq. (\ref{Icoll}) local off-shell transition
amplitudes enter for the transitions $P + P_2 \rightarrow P_3 +
P_4$. We note that the extension of Eq. (\ref{Icoll}) to inelastic
scattering processes (e.g. $NN \rightarrow N\Delta$) or ($\pi N
\rightarrow \Delta$ etc.) is straightforward when exchanging the
elastic transition amplitude $T$ by the corresponding inelastic
one and taking care of Pauli-blocking or Bose-enhancement for the
particles in the final state. The relation of the quantity
$\Gamma_{XP}$ to the collisional width $\Gamma_{coll}$ is given by
$\Gamma_{XP} = 2 P_0 (\Gamma_{decay}(XP) + \Gamma_{coll}(XP))$,
where the particle decay width $\Gamma_{decay}$ in the medium
might also be different compared to the vacuum.

Thus the transport approach determines the particle spectral
function dynamically via (\ref{gcoll}) -- with respect to the collisional
width $\Gamma_{coll}$ -- for all hadrons if the
in-medium transition amplitudes $T$ are known in their full
off-shell dependence. Since this information is not available for
configurations of hot and dense matter, a couple of assumptions
and numerical approximation schemes have to be invoked in actual
applications so far.

As in Refs. \cite{ca1,ca2} the following dynamical calculations are
based on the conventional HSD transport approach
\cite{CB99,Ehehalt} -- in which $Re \Sigma^{ret}_{XP}$ is
specified for the hadrons -- however, the equations of motion for
the testparticles are extended to (\ref{eomr},\ref{eomp},\ref{eomm}).
For further details on the elastic and inelastic transition rates we
refer the reader to Ref. \cite{ca2}.

\subsection{Comment on particle number conservation}
In previous derivations of the off-shell transport equations
one has started from a formulation of the non-equilibrium theory
in space-time representation $(x,x')$ and then changed into a
phase-space representation via Fourier transformation with respect
to $(x-x')$ \cite{Bot,ca1,ca2,Leupold}.
The semiclassical limit then has been introduced by assuming gradients
in $X=(x+x')/2$ to be small \cite{Leupold} for $Re \Sigma^{ret}_{XP}$ and $S^<_{XP}$.
Here we argue that for reasons of symmetry in phase-space also the
four-momentum derivatives in $P$ have to be small to achieve a
proper semiclassical limit as inherent in the formulation of the
transport equation (\ref{trans_approx})  in
terms of the generalized Poisson-bracket (\ref{poissonoperator}).

We briefly demonstrate in the following lines that instead of a coordinate-space
representation one may formulate the theory equivalently in
momentum-space and then transform to phase space:
The momentum-dependent (two-point) functions are given as
\bea F_{p_{1\php} \, p_{1'}} \: = \:
\int \! d^{4}\!x_1 \! \int d^{4}\!x_{1'} \; \;
e^{i ( p_{1\php}x_{1\php} - \, p_{1'} x_{1'})}
\; \; F_{x_{1\php} \, x_{1'}}.
\label{fourier_transformation}
\eea\\
The evolution of the retarded and advanced Green functions $S^{ret},
S^{adv}$ and the Green function $S^<$ turns to
\bea
\hat{S}_{0 \, p_{1\php}}^{-1} \; S_{p_{1\php} \, p_{1'}}^{ret,adv}
\; & = & \:
(2 \pi)^4 \; \delta^{(4)}(p_{1\php} - p_{1'})
\; \: +\: \:
\int \frac{d^{4}p_2 }{(2 \pi)^{4}} \; \;
\Sigma_{p_{1\php} \, p_{2\php}}^{ret,adv} \; \;
S_{p_{2\php} \, p_{1'}}^{ret,adv} \; , \\
\hat{S}_{0 \, p_{1\php}}^{-1} \; S_{p_{1\php} \, p_{1'}}^{<}
\; & = & \:
\int \frac{d^{4}p_2 }{(2 \pi)^{4}} \; \; \left[ \;
\Sigma_{p_{1\php} \, p_{2\php}}^{ret} \; S_{p_{2\php} \, p_{1'}}^{<}
\; \: + \; \:
\Sigma_{p_{1\php} \, p_{2\php}}^{<} \; \ S_{p_{2\php} \, p_{1'}}^{adv}
\; \right] \: ,
\label{kb_momentum}
\eea \\
with the 'kinetic' operator in momentum-space
$\hat{S}_{0 \, p_1}^{-1} = (p_1^2-M^2_0)$ in the case of
relativistic scalar bosons with (bare) mass $M_0$.
Obviously, the equations in momentum-space are formally equivalent
to the equations in coordinate-space, i.e. the convolution integrals
in the coordinate $x_2$ are replaced by convolution integrals in
the momentum $p_2$.
When transforming from momentum- to phase-space via
a Fourier-transformation with respect to the four-momentum
coordinate $(p_1-p_{1'})$
\bea
F_{XP} \: = \: \int \frac{d^{4}(p_1\!-\!p_{1'})}{(2 \pi)^4} \; \;
e^{- i X (p_{1\php} - p_{1'})} \; \; F_{p_{1\php} \, p_{1'}}
\label{wigner_transformation_2}
\eea\\
(with $P=(p_1+p_{1'})/2$) we gain again the familiar equations in
Wigner-representation:
\bea
\hat{S}_{0 \, XP}^{-1} \; S_{XP}^{ret,adv}
\; & = & \; 1
\; \: + \: \:
e^{- i \Diamond} \; \; \Sigma_{XP}^{ret,adv} \; S_{XP}^{ret,adv} \\[0.5cm]
\hat{S}_{0 \, XP}^{-1} \; S_{XP}^{<}
\; & = & \:
e^{- i \Diamond} \;
\left[ \; \Sigma_{XP}^{ret} \; \;
S_{XP}^{<}
\; \: + \; \:
\Sigma_{XP}^{<} \; \;
S_{XP}^{adv} \; \right] \: ,
\label{kb_wigner}
\eea \\
with the 'kinetic' operator in phase-space
$\hat{S}_{0 \, XP}^{-1} = (P^2 + i P^{\mu} \partial^{X}_{\mu}
- \frac{1}{4} \partial^{X}_{\mu} \partial_{X}^{\mu} - M^2_0)$.
Here we have used the following relation for the convolution
integrals in momentum-space
\bea
\int \frac{d^{4}(p_1\!-\!p_{1'})}{(2 \pi)^4} \; \;
e^{-i X (p_{1} - p_{1'})} \;
\int \frac{d^{4}p_2}{(2 \pi)^4} \; \;
\; F_{1, \, p_{1\php} \, p_{2\php}} \; F_{2, \, p_{2\php} \, p_{1'}}
\; \; = \; \; e^{-i \Diamond} \; \; F_{1, \, PX} \; \;
F_{2, \, PX},
\label{wigner_convolution}
\eea\\
with the Poisson-operator $\Diamond$ defined in (\ref{poissonoperator}).

The semiclassical limit -- along the conventional line of arguments --
now is achieved by assuming gradients of the self energies $\Sigma^{ret,adv}$ in
$P$ to be small. This assumption is especially well taken for systems
with dominantly short range interactions since the different self energies
$\Sigma^{ret,adv}(p,p')$ become smooth functions in momentum-space
such that a restriction to first order gradients in the momentum
$P=(p+p')/2$ can be more easily justified. Note again that
the four-dimensional Poisson-bracket (\ref{poissonoperator}) is
symmetric in the space-time and the four-momentum derivatives.
Thus a phase-space gradient expansion requires all gradients
to be small contrary to the assumptions made in Ref. \cite{Leupold}.

As a consequence mixed gradients of second order as
$\partial^2/(\partial t \partial p_0)$
have to be neglected in a consistent truncation scheme of first
order. As shown by Botermans and Malfliet in the review \cite{Bot} the particle
number conservation then holds strictly.
Only when keeping special terms of second order
$(\sim \partial^2/(\partial t \partial p_0)$
unphysical pecularities may appear, that lead to a violation of particle
number conservation in contradiction to the full theory (2)-(4).

\section{Infinite nuclear matter problems}
In case of infinite nuclear matter problems the solution of the
transport equations simplifies since all spatial gradients
(with respect to $\vec{X}$) vanish. This implies that the momentum
coordinates of the testparticles $\vec{P}_i$ are constant
according to (19) in between collisions.

The initial conditions of the problem then are fully specified by an initial
bombarding energy per nucleon, that defines the relative shift in
momentum of two Fermi spheres with a Fermi momentum $P_F$ = 260 MeV/c,
the number of nucleons $N$ and the total volume of the box
$V=L^3$.  Alternatively, one can also characterize the system by an
average density $\rho =N/V$ and energy density $\epsilon$ (cf.
Ref. \cite{Brat2000}). We use $L = 10$ fm and assume an equal
number of neutrons and protons.

In general (for $t \rightarrow \infty$) the stationary solution of
Eq. (\ref{Icoll}) for a fermion $h$ is given by
\begin{equation}
\label{equilibrium}
F_h(\vec{X}, \vec{P},M^2)
\; = \; \frac{A_h({\vec{X}},{\vec{P}},M^2)}{\exp((E-\mu_h)/T)-\eta}
\end{equation}
with $E=\sqrt{{\vec{P}}^2 + M^2}$ and $\eta = \pm 1$ for bosons/fermions,
respectively,
while $A_h$ denotes the spectral function (\ref{alg_spectral}),
$T$ the temperature of the system and $\mu_h$ the chemical
potential for the hadron. This situation corresponds to the grand canonical
ensemble of quantum statistical mechanics. In case of infinite nuclear matter
problems the dependence on $\vec{X}$ vanishes additionally.

\subsection{Numerical results}
\begin{figure}[h]
{\vspace*{-1.0 cm} \hspace*{0.0 cm}
{\epsfig{file=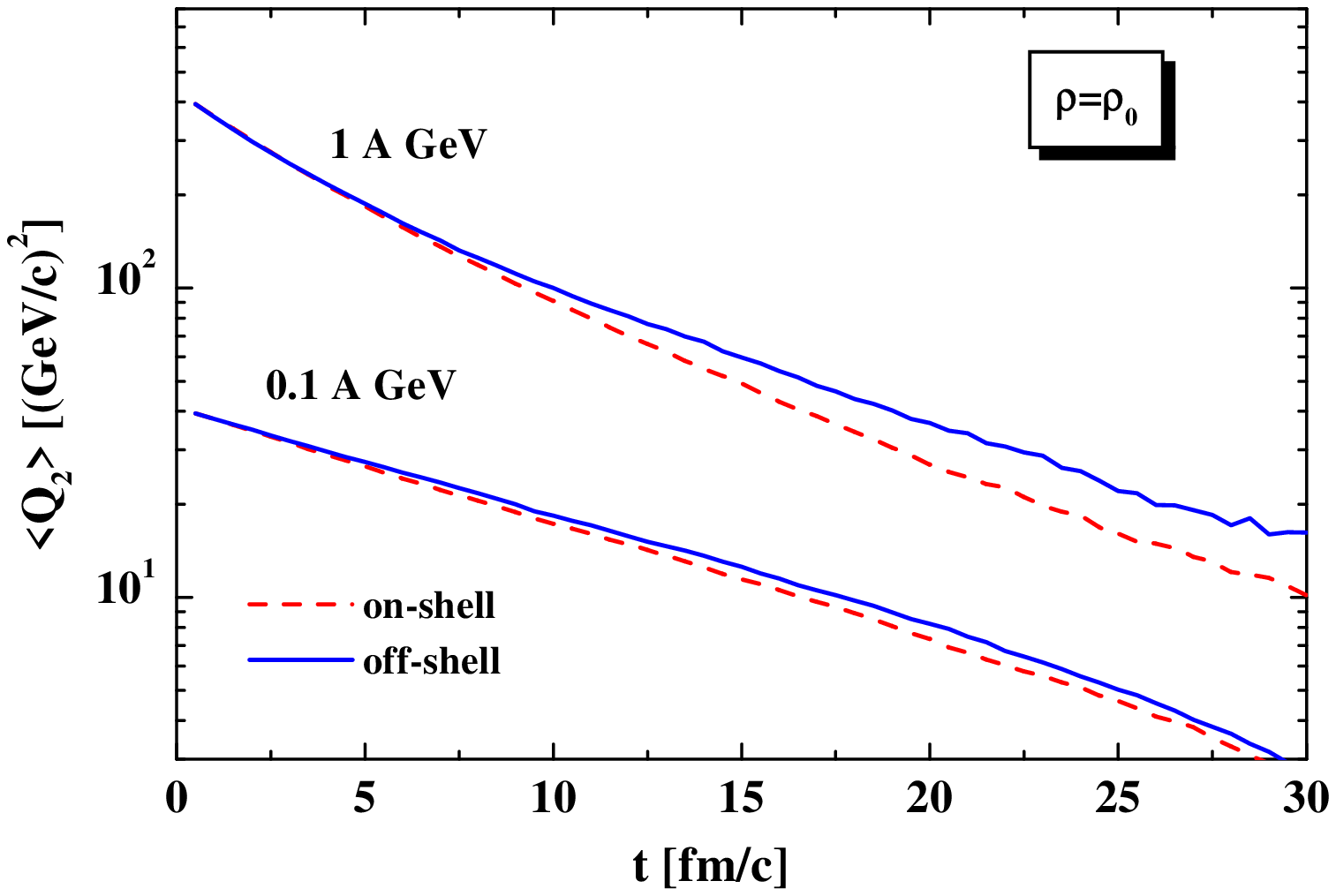,height=20 cm,width=15 cm}}} \caption{The
quadrupole moment in momentum space (\protect\ref{quad}) for an
infinite nuclear matter problem in a finite box with periodic
boundary conditions characterized by  bombarding energies of 0.1
A GeV and 1 A GeV at density $\rho = \rho_0$. The solid lines
present the results for the off-shell calculations while the
dashed lines are obtained in the on-shell limit.
 \label{bild1}}
\end{figure}

As an example for equilibration phenomena we show in Fig.
\ref{bild1} the time evolution of the quadrupole moment (involving
all baryons $B$)
\be
\label{quad} Q_2(t) = \sum_B \frac{g_B}{(2 \pi)^4} \int d^3 X \int d^3 P
\, \frac{d M^2}{2 \sqrt{{\vec P}^2 + M^2}} \ (2P_z^2-P_x^2-P_y^2)\
F_B(\vec{X},t,\vec{P},M^2), \ee for $\rho=\rho_0$ and initial
bombarding energies of 0.1 A GeV and 1 A GeV, respectively.
In (\ref{quad}) the degeneracy factor is $g_B=4$ for nucleons and $g_B$= 16
for $\Delta$-resonances. The
solid lines result from the off-shell calculation  while the
dashed lines result from the on-shell limit. As can be seen from
Fig. \ref{bild1} the off-shell results are practically identical
to the on-shell limits except for the very long time behaviour, where the
off-shell limit needs some more time to achieve equilibrium.
However, when defining an equilibration time $\tau$
by a drop of the observable by the factor $e^{-1}$ we find no
sizeable effect from the off-shell propagation on $\tau$.

\begin{figure}[h]
{\vspace*{-1.0 cm} \hspace*{0.0 cm} {\epsfig{file=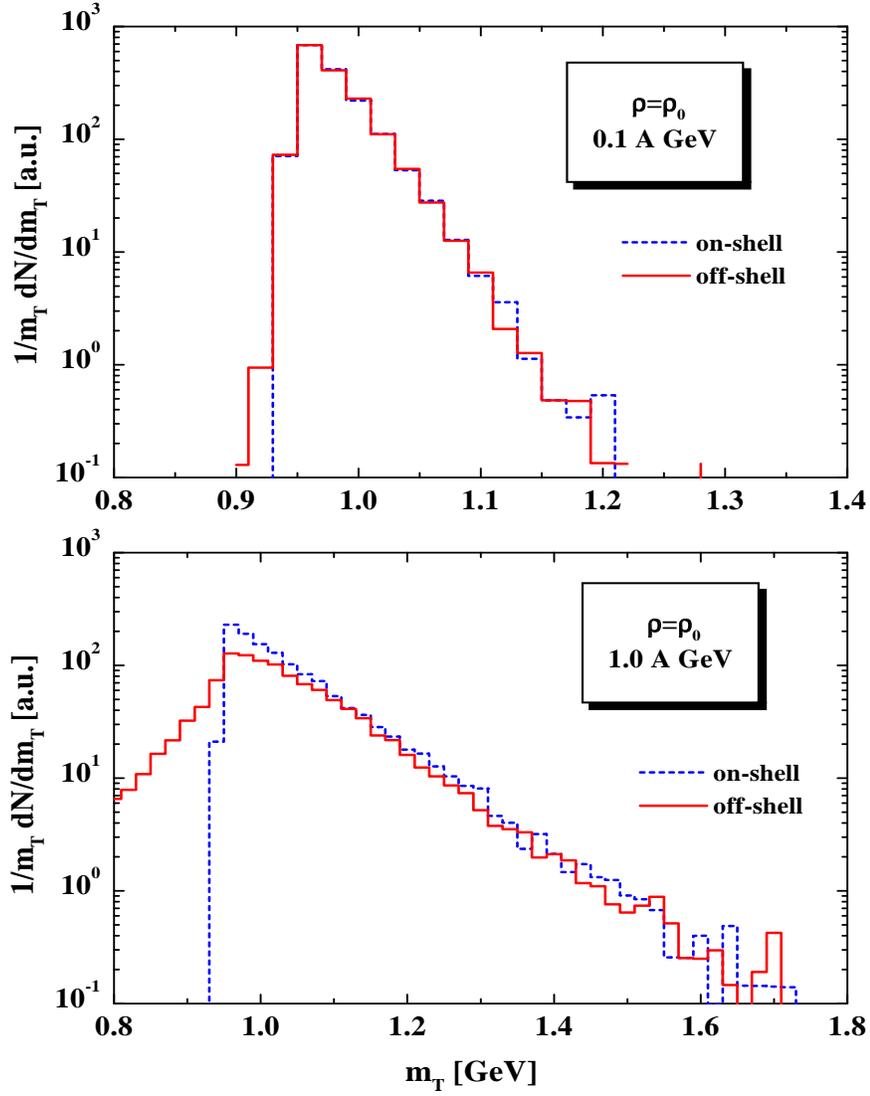,height=20
cm,width=15 cm}}} \caption{The transverse mass distribution for an
infinite nuclear matter problem at bombarding energies of 0.1 A
GeV (upper part) and 1 A GeV (lower part) at density $\rho =
\rho_0$. The solid lines present the results for the off-shell
calculations while the dashed lines are obtained in the on-shell
limit.
 \label{bild2}}
\end{figure}

The equilibrium distributions in the nucleon transverse mass $M_t
= \sqrt{p_t^2 + M^2}$ are shown in Fig. \ref{bild2} for initial
bombarding energies of 0.1 A GeV (upper part) and 1 A GeV (lower
part). Again the solid lines denote the off-shell results from the
transport calculations while the on-shell spectra are displayed in
terms of dashed lines. Both limits (within the statistics)
give the same temperature $T \approx$ 97 MeV for 1.0 A GeV and $T
\approx$ 26 MeV for 0.1 A GeV as can be extracted from the high
energy tail of the transverse mass spectra. Differences can only
be found for $m_T \leq M_0$ since the finite width in the nucleon
spectral function only can show up in the off-shell case. This
component is quite small at 0.1 A GeV since the collisional width
of nucleons at density $\rho_0$ and temperature $T \approx$ 26 MeV
is about 8 MeV.

Without explicit display we mention that the time evolution of the
nucleon spectral function in the invariant mass $M$ becomes broad in
the initial nonequilibrium phase of the reaction and approaches the
equilibrium distribution (\ref{equilibrium}) roughly within the
equilibration times from Fig. \ref{bild1}. Since the width of the
nucleon spectral function in equilibrium at 0.1 A GeV (GANIL
energy) is rather small ($\Gamma \approx $ 8 MeV) we concentrate on
the energy of 1 A GeV (SIS energy) in the following, where the
nucleon collisional width $\Gamma_{coll}$ at a temperature of 97
MeV is about 40 MeV and roughly 20\% of the baryons are excited
$\Delta$ resonances\footnote{Note, that in collisions of finite systems
at this bombarding energy the $\Delta$-abundancy is slightly lower due
to a rapid expansion of the system and the additional compressional energy
stored in the system in the high density phase.}.

As demonstrated above, for the equilibrated system we can extract
a temperature $T$ by
fitting the particle spectra with the Bolzmann distribution
\begin{eqnarray}
{d^3N_i\over dp^3} \sim \exp(-E_i/T), \label{Boltz}\end{eqnarray}
where $E_i=\sqrt{p_i^2+m_i^2}$ is the energy of particle $i$.  We
note that at the temperatures of interest here, the Bose and Fermi
distributions are practically identical to a Boltzmann
distribution. We find that in equilibrium  the spectra of all
particles (nucleons, $\Delta$'s and pions) can be characterized by
a single temperature $T$ (cf. e.g. Ref. \cite{Brat2000}).

\subsection{Comparison to the statistical model}
In order to investigate the equilibrium behaviour of hadron matter
we compare our transport (box) calculations with a simple
Statistical Model (SM) for an Ideal Hadron Gas where the
system is described by a grand canonical ensemble of
non-interacting fermions and bosons in equilibrium at temperature
$T$.  All baryon and meson species considered in the transport
model ($N, \Delta, \pi$) also are included in the
statistical model.

We recall that in  the SM particle multiplicities
$n_i$ and energy densities $\varepsilon_i$ for particles with
spectral functions $A_i$ are given by
\begin{eqnarray}
&& n_i ={g_i \over (2\pi \hbar)^3} \int \frac{dM}{2 \pi}
\int\limits_0^\infty {A_i(M)/ 4\pi p^2 dp \over \exp\left[(E_i -
B_i\mu_B - S_i \mu_S)/T\right] - \eta}, \label{Nth} \\ &&
\varepsilon_i ={g_i \over (2\pi \hbar)^3} \int \frac{dM}{2 \pi}
 \int\limits_0^\infty {A_i(M)\ 4\pi E_i p^2 dp \over \exp\left[(E_i
- B_i\mu_B - S_i\mu_S)/T\right]- \eta}, \label{Enth}
\end{eqnarray}
where $E_i = \sqrt{p^2+M^2}$ is the energy of particle $i$, $B_i$
is the baryon charge, $S_i$ is the strangeness, and $g_i$ is the
spin-isospin degeneracy factor, while $\eta=\pm 1$ for
bosons/fermions, respectively. In Eqs. (\ref{Nth}),(\ref{Enth})
$\mu_B$ and $\mu_S$ are the baryon and strangeness chemical
potentials. The energy density $\varepsilon$, baryon density
$\rho$ and strange particle density of the whole system in
equilibrium then is given as
\begin{eqnarray}
&& \varepsilon =  \sum\limits_i \varepsilon_i (T,\mu_B,\mu_S)
\label{3eq_en} \\ && \rho = \sum\limits_i B_i \ n_i
(T,\mu_B,\mu_S) \label{3eq_rhoB} \\ && \rho_S = \sum\limits_i S_i
\ n_i (T,\mu_B,\mu_S) \, \equiv \, 0 . \label{3eq_rhoS}
\end{eqnarray}
 As 'input' for the SM we use the same $\varepsilon, \rho$
and $\rho_S$ as in the box calculations and  we obtain the
thermodynamical parameters -- $T, \mu_B, \mu_S$ -- by solving the
system of nonlinear equations (\ref{3eq_en}),(\ref{3eq_rhoB}) and
(\ref{3eq_rhoS}).

We now turn to a comparison of the equilibrium distributions in
mass for nucleons and $\Delta$'s, i.e. $dN_N/dM$ and
$dN_\Delta/dM$, from the box calculations with those from the SM,
which are obtained by integration (summation) over momentum. We
first discuss the on-shell limit where the nucleon spectral
function is represented by a $\delta$-function around the bare
mass while the $\Delta$ spectral function -- as implemented in the
transport approach -- is given by (in the $\Delta$ rest frame)
\begin{equation}
\label{delta} A_\Delta(M) \; = \; \frac{ 2 M^2 \Gamma_{\pi
N}(M)}{(M^2-M_{\Delta0}^2)^2 + M^2 \Gamma^2_{\pi N}(M)}
\end{equation}
with
\begin{equation}
\label{delta2} \Gamma_{\pi N}(M) = \Gamma_0 (\frac{q}{q_R})^3 \
(\frac{q_R^2 + \delta^2} {q^2 + \delta^2})^3 ,
\end{equation}
where $q_R$ denotes the pion momentum in the $\Delta$ rest frame
of the resonance and $q$ is the corresponding pion three-momentum
at invariant mass $M$, i.e.
\begin{equation}
q^2 = \frac{(M^2 - (M_N+M_\pi)^2) (M^2 - (M_N-M_\pi)^2)}{4 M^2}.
\end{equation}
The quantity $\delta$ in (\ref{delta2}) is fixed by \cite{Teis}
\begin{equation}
\delta^2 = (M - M_N - M_\pi)^2 + \frac{\Gamma_0^2}{4}
\end{equation}
with $\Gamma_0 = 120$ MeV to achieve a good
description for the $\pi N \rightarrow \Delta$
reaction in vacuum. Recall that in this case we have
$\Gamma_{XP}/(2 M) = \Gamma_{decay} = \Gamma_{\pi N}$ in the
$\Delta$ rest frame, i.e. $\Gamma_{coll}$ = 0.

\begin{figure}[h]
{\vspace*{-3.0 cm} \hspace*{0.0 cm}
{\epsfig{file=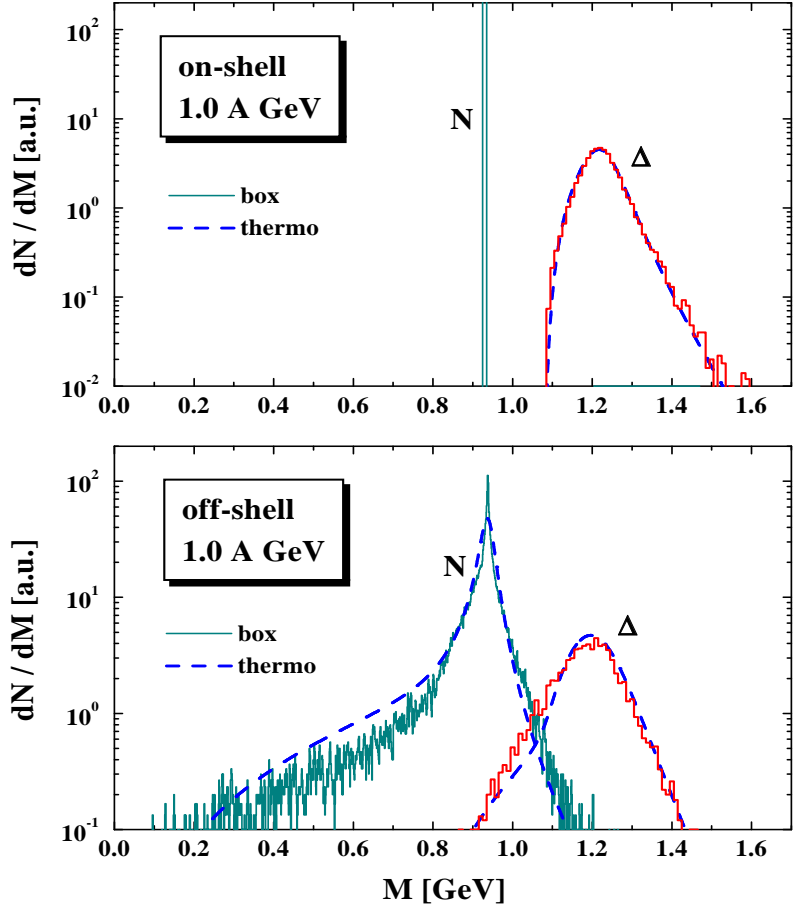,height=20 cm,width=15 cm}}} \caption{The
differential distribution in mass for nucleons and $\Delta$
resonances at equilibrium for an infinite nuclear matter problem
at an initial  bombarding energy of 1 A GeV at density $\rho =
\rho_0$. The solid histograms present the results from the
transport (box) calculations for the on-shell limit (upper part)
and off-shell limit (lower part), respectively. The dashed lines
('thermo') are obtained from the statistical model at temperature $T= $ 97
MeV employing the spectral functions from the transport approach
(see text).
 \label{bild3}}
\end{figure}

In the upper part of Fig. \ref{bild3} we compare the asymptotic ($t
\rightarrow \infty$)  distributions for nucleons ($N$) and
$\Delta$'s in the on-shell limit from the transport (box)
calculation (solid histograms) with the result from the SM at a
temperature $T$ = 97 MeV employing the $\Delta$ spectral function
(\ref{delta}). Since the differences are hardly visible, we
conclude that the transport calculation reproduces the result from
the SM, where the thermodynamical parameters are determined by
energy and baryon number conservation. We mention that we have
discarded strangeness in this comparison, since kaons and hyperons
are very scarce at this energy and have been switched off in both
models. Since the $\Delta$ width (\ref{delta2}) is zero below the
$\pi N$ threshold, the $\Delta$ mass distribution only can extend
above $M_{\pi} + M_N$.

In case of the off-shell calculation we obtain somewhat different
mass distributions for nucleons and $\Delta$'s from the box
calculation, which are given in terms of the solid histograms in
the lower part of Fig. \ref{bild3}. Here the nucleon spectral
function becomes very broad and also the $\Delta$ distribution
extends below the $\pi N$ threshold in vacuum. The nucleon
spectral function is broadened due to the elastic ($NN \rightarrow
NN$) and inelastic ($NN \rightarrow N\Delta$) scattering
processes, which can roughly be described by a collisional width
$\Gamma_{coll}$ of 40 MeV in the nucleon spectral function
\begin{equation}
\label{nucleon} A_N(M) \; = \; \frac{2 M^2 \Gamma_{coll}}
{(M^2-M_{N0}^2)^2 + M^2 \Gamma_{coll}^2},
\end{equation}
as shown by the left dashed line in the lower part of Fig.
\ref{bild3}. The collisional width in case of nucleons is related to $\Gamma_{XP}$
as $2P_0 \Gamma_{coll}=\Gamma_{XP}$ which reduces in the rest
system of the particle to $2M \Gamma_{coll}=\Gamma_{XP}$.

The $\Delta$ mass spectrum in this case is more difficult to
interprete. This is due to the fact that in the decay $\Delta
\rightarrow \pi N^*$ the $\Delta$ may decay to a pion and an
off-shell nucleon $N^*$ according to the nucleon equilibrium
spectral function $A_N(M)$, which essentially shifts the $\pi N^*$
threshold accordingly. However, taking into account this change in
width due to the in-medium $\pi N^*$ decay, only, the low mass
part of the $\Delta$ distribution is underestimated considerably.
Here the 'collisional' channels $\Delta N \rightarrow \Delta N$
and $\Delta N \rightarrow NN$ are much more important. We mention
that the latter reaction is described by the extended detailed
balance relation of Ref. \cite{Wolf} while the elastic
differential cross section is taken the same as for $NN$
scattering (as a function of the momentum transfer).  Since
especially the $\Delta N$ absorption reaction depends on mass $M$
and momentum $p$ of the baryons explicitly, it is not
straightforward to present analytical formulas since final state
Pauli blocking for the nucleons leads to a highly nonlinear
problem. We thus have extracted the $\Delta$ collisional width
$\Gamma^\Delta_{coll}(p,M)$ from the transport calculation
explicitly by calculating the number of $\Delta N$ unblocked
collisions per unit time as a function of the $\Delta$ momentum
$p$ and mass $M$. Using ($p = |\vec{P}|$)
\begin{equation}
\label{delta3} \frac{\Gamma_{XP}}{2 P_0} = \Gamma_{tot}(p,M) = \Gamma_{\pi N^*}(M) +
\Gamma^\Delta_{coll}(p,M) ,
\end{equation}
where $\Gamma_{\pi N^*}$ denotes the $\Delta$ in-medium $\pi N^*$
width averaged over the nucleon equilibrium distribution function
(cf. lower part of Fig. 3) and integrating over momentum with the
appropriate Fermi function (for $T$ = 97 MeV) we obtain the right
dashed line in the lower part of Fig. \ref{bild3} that describes
the box $\Delta$- distribution rather well. Thus the transport
calculation at finite density $\rho_0$ and temperature $T$ is
consistent with the SM when employing the same physical baryon
spectral functions. One might have expected this equivalence in
equilibrium due to energy and baryon number conservation, however,
the actual numerical result then may be regarded as a consistency
test of the numerical implementation schemes for the various
elastic and inelastic reaction channel in the medium for off-shell
particle propagation.

\begin{figure}[h]
{\vspace*{-1.0 cm} \hspace*{1.5 cm}
{\epsfig{file=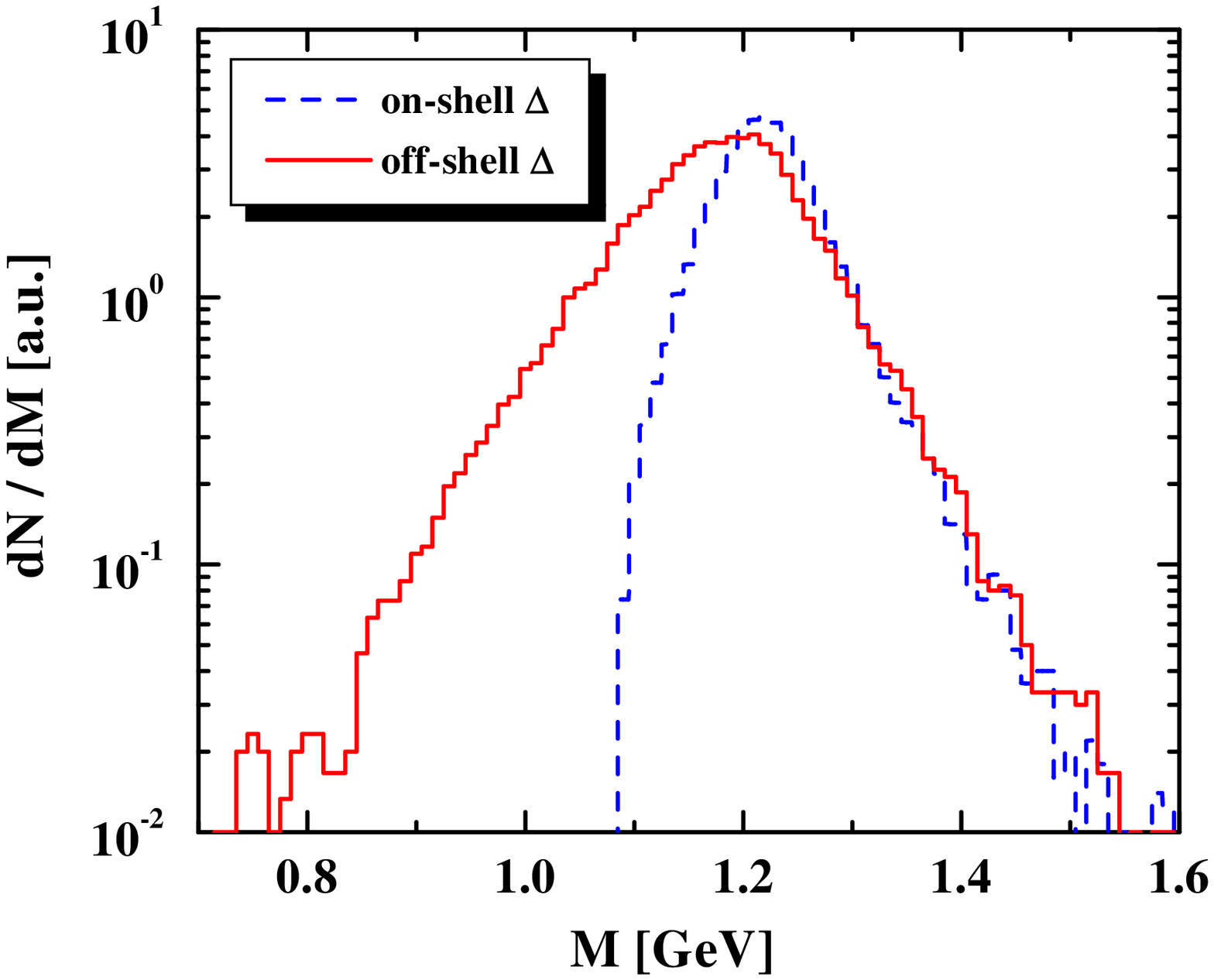,height=15 cm,width=12 cm}}}
\caption{Comparison of the differential distribution in mass for
 $\Delta$ resonances at equilibrium for the same infinite
nuclear matter problem as in Fig. 3. The solid histogram presents
the results from the transport  calculations for the off-shell
limit while the dashed histogram stems from the on-shell
calculation.
 \label{bild4}}
\end{figure}

The mass integrated number of $\Delta$'s in the off-shell case is
larger by a few \% as compared to the on-shell case in equilibrium
due to the low mass tail of the $\Delta$ distribution, however,
the high mass tails of the $\Delta$'s are identical within the
statistics achieved as demonstrated in Fig. \ref{bild4}. On the
other hand, the number of pions is practically identical in both
cases since the low mass tail of the $\Delta$'s is dominantly due
to $\Delta N$ reactions and only to a lower extent to the $\pi
N^*$ decay as discussed above.

\section{Summary}
In this work we have employed the semiclassical off-shell approach
from Refs. \cite{ca1,ca2} to study equilibration phenomena in
intermediate and high energy nucleus-nucleus collisions in a
finite box with periodic boundary conditions. The semiclassical
off-shell transport approach describes the virtual propagation of
particles in the invariant mass squared $M^2$ besides the
conventional propagation in the mean-field potential given by the
real part of the self energy. The imaginary part of the retarded
self energy $Im \Sigma^{ret}_{XP} = -1/2 \Gamma_{XP} = - P_0
(\Gamma_{coll}(XP)
+ \Gamma_{decay}(XP))$  -- apart from decay contributions to the
width  -- is determined by the collision integrals
themselves and 'evaluated' within the transport approach dynamically.

Our explicit calculations demonstrate that the off-shell
propagation of nucleons has practically no sizeable effect on
equilibration times especially at lower bombarding energies; more
importantly, the off-shell dynamics even lead to a slight increase
of the equilibration time for kinetic equilibrium as noticed early by
Danielewicz \cite{pd841} (cf. also Ref. \cite{CGreiner}).
Furthermore, we have demonstrated that the off-shell HSD approach
reproduces the 'proper' spectral functions with respect to the
statistical model (in case of a grand canonical ensemble) for nucleons and
$\Delta$'s that in equilibrium are given analytically once the
collisional width $\Gamma_{coll}(\vec{P}, M)$ is known as a
function of the 3-momentum $\vec{P}$ and invariant mass $M$.

\vspace{1cm} \noindent The authors like to acknowledge stimulating
discussions with E.L. Bratkovskaya, C. Greiner and S. Leupold
throughout this study.

%----------------------------------------------------------------------

%
%
%----------------------------------------------------------------------


\begin{thebibliography}{999}
\bibitem{URQMD}
    S. Bass, M. Belkacem, M. Bleicher et al.,
    Prog. Part. Nucl. Phys. 41 (1998) 255.
\bibitem{CB99}
    W. Cassing and E. L. Bratkovskaya, Phys. Rep. 308 (1999) 65.
\bibitem{kb62} L. P. Kadanoff and G. Baym,
    {\it Quantum statistical mechanics}, Benjamin, New York, 1962.
\bibitem{pd841}
   P. Danielewicz, Ann. Phys. (N.Y.) 152 (1984) 239; {\it ibid.} 305.
\bibitem{Bot}
    W. Botermans and R. Malfliet, Phys. Rep. 198 (1990) 115.
\bibitem{Mal}
    R. Malfliet, Prog. Part. Nucl. Phys. 21 (1988) 207.
\bibitem{ph95}
    P. A. Henning, Nucl. Phys. A 582 (1995) 633; Phys. Rep. 253 (1995) 235.
\bibitem{gl98}
    C. Greiner and S. Leupold, Ann. Phys. (N.Y.) 270 (1998) 328.
\bibitem{Zuo}
    S. J. Wang, W. Zuo and W. Cassing, Nucl. Phys. A 573 (1994) 245.
\bibitem{Wang1}
  S. J. Wang and W. Cassing, Ann. Phys. (N.Y.) 159 (1985) 328.
\bibitem{CaWa}
   W. Cassing and S. J. Wang, Z. Phys. A 337 (1990) 1.
\bibitem{CNW}
    W. Cassing, K. Niita and S. J. Wang, Z. Phys. A 331 (1988) 439.
\bibitem{Rudy1}
    R. Malfliet, Nucl. Phys. A 545 (1992) 3.
\bibitem{Rudy2}
    R. Malfliet, Phys. Rev. B 57 (1998) R11027.
\bibitem{ca1} W. Cassing and S. Juchem, Nucl.
Phys. A 665 (2000) 385.
\bibitem{ca2} W. Cassing and S. Juchem, nucl-th/9910052, Nucl.
Phys. A, in print.
\bibitem{Leupold} S. Leupold, nucl-th/9909080 Nucl. Phys. A, in print.
\bibitem{Brat2000}
  E. L. Bratkovskaya, W. Cassing, C. Greiner et al., nucl-th/0001008.
\bibitem{Brav1}
    M. Belkacem, M. Brandstetter, S.A. Bass et al.,
    Phys. Rev. C 58 (1998) 1727.
\bibitem{Brav2}
    L.V. Bravina, M.I. Gorenstein, M. Belkacem et al.,
    Phys. Lett. B 434 (1998) 379;
    L.V. Bravina, M. Brandstetter, M.I. Gorenstein et al.,
    J. Phys. G 25 (1999) 351.
\bibitem{Brav3}
    L.V. Bravina, E.E. Zabrodin, M.I. Gorenstein et al.,
    Phys. Rev. C 60 (1999) 024904.
\bibitem{Solfr99}
    J. Sollfrank, U. Heinz, H. Sorge, N. Xu, Phys. Rev. C 59 (1999) 1637.
\bibitem{Effe1} M. Effenberger, E. L. Bratkovskaya and U. Mosel,
Phys. Rev. C 60 (1999) 44614.
\bibitem{Effe} M. Effenberger and U. Mosel, Phys. Rev. C 60 (1999) 51901.
\bibitem{Knoll} Yu. B. Ivanov, J. Knoll, and D. N. Voskresensky, nucl-th/9905028.
\bibitem{Ehehalt}
    W. Ehehalt and W. Cassing, Nucl. Phys. A 602 (1996) 449.
\bibitem{Teis} S. Teis, W. Cassing, M. Effenberger et al., Z. Phys.  A356 (1997) 421.
\bibitem{Wolf} Gy. Wolf, W. Cassing and U. Mosel, Nucl. Phys. A 552 (1993) 549.
\bibitem{CGreiner} C. Greiner, K. Wagner and P.G. Reinhard,
                   Phys. Rev. C 49 (1994) 1693.
\end{thebibliography}
\end{document}